\documentstyle[11pt,aaspp4]{article}

\newcommand {\hii}{H\,{\sc ii}}
\newcommand {\hi}{H\,{\sc i}}
\newcommand {\ha}{H$\alpha$}

\newcommand {\rah}{$^{\rm h}$}
\newcommand {\ram}{$^{\rm m}$}

\newcommand {\kms}{km~s$^{-1}$}
\newcommand {\rosat}{{\it ROSAT\/}}
\newcommand {\asca}{{\it ASCA\/}}
\newcommand {\einst}{{\it Einstein\/}}
 
\newcommand {\lmc}{LMC\,2}

\newcommand {\him}{10$^6$~K}

\newcommand{\Msun}{$M_{\odot}$}
\newcommand{\csam}{counts~s$^{-1}$~arcmin$^{-2}$}
 
\begin{document}

\title{The Supergiant Shell LMC\,2: II. Physical Properties of the \him\ Gas}

\author{S. D. Points, Y.-H. Chu}

\affil{Astronomy Department, University of Illinois 
       1002 W. Green St., Urbana, IL 61801, USA}
 
\author{S. L. Snowden\altaffilmark{1}}
\affil{NASA/Goddard Space Flight Center 
       Code 662, Greenbelt, MD 20771, USA}

\author{L. Staveley-Smith}
\affil{Australia Telescope National Facility, CSIRO, P.O. Box 76, Epping,
NSW 2121, Australia}

\altaffiltext{1}{Universities Space Research Association} 

\begin{abstract}

\lmc\ has the highest X-ray surface brightness of all know supergiant 
shells in the Large Magellanic Cloud (LMC).  The X-ray emission peaks 
within the ionized filaments that define the shell boundary, but also 
extends beyond the southern border of \lmc\ as an X-ray bright spur.  
\rosat\ HRI images reveal the X-ray emission from \lmc\ and the spur 
to be truly diffuse, indicating a hot plasma origin.  We have obtained 
\rosat\ PSPC and \asca\ SIS spectra to study the physical conditions of 
the hot ($\ge 10^6$~K) gas interior to \lmc\ and the spur.  Raymond-Smith 
thermal plasma model fits to the X-ray spectra, constrained by \hi\ 
21-cm emission-line measurements of the column density, show the plasma 
temperature of the hot gas interior of \lmc\ to be $kT \sim 0.1 - 0.7$~keV 
and of the spur to be $kT \sim 0.1 - 0.5$~keV.  We have compared the 
physical conditions of the hot gas interior to \lmc\ with those of other
supergiant shells, superbubbles, and supernova remnants (SNRs) in the 
LMC.  We find that our derived electron densities for the hot gas inside 
\lmc\ is higher than the value determined for the supergiant shell
LMC\,4, comparable to the value determined for the superbubble N\,11,
and lower than the values determined for the superbubble N\,44 and 
a number of SNRs.

\end{abstract}

\keywords{ISM: bubbles -- ISM: individual (LMC\,2) -- X-rays: interstellar
X-rays: diffuse emission}

\clearpage

\section{Introduction}

Supergiant shells with sizes approaching 1000~pc form the largest
structures seen in the interstellar medium (ISM).  Such shells have
been detected in the \hi\ 21-cm and \ha\ lines in our Galaxy, Local 
Group galaxies, and more distant galaxies (Tenorio-Tagle \& Bodenheimer 
1988).  These supergiant shells contain 10$^{51}$--10$^{54}$ ergs 
of kinetic energy, which is most likely contributed by multiple 
supernovae and energetic stellar winds over several dozen Myr.  The 
size of a supergiant shell often exceeds the scale height of the 
gaseous disk of its host galaxy, thus forming a ``chimney'' through 
which energy and mass flow into the galactic halo.  Therefore, 
supergiant shells play important roles in the global evolution of 
the ISM in a galaxy.

It is difficult to study supergiant shells in detail. Their 
distances, and so their sizes and energies, remain uncertain in 
our Galaxy; furthermore, interstellar obscuration often prevents
observations at optical, UV and X-ray wavelengths.  For distant
galaxies, angular resolution becomes a severe limiting factor.
The Large Magellanic Cloud (LMC) offers a happy compromise, as it
is close (50~kpc; Feast 1999) and has little obscuration (A$_V 
< 0.3$ mag; Bessell 1991)

Nine supergiant shells with sizes ranging from 600 to 1400~pc have
been reported in the LMC, LMC\,1--9 (Goudis \& Meaburn 1978; Meaburn 
1980).  Only two of these shells have been studied in detail -- 
LMC\,2 and LMC\,4 -- because LMC\,2 is the most spectacular and LMC\,4 
is the largest.  Both supergiant shells are filled with hot plasma: 
diffuse X-ray emission has been detected in LMC\,2 by Wang \& Helfand 
(1991) using \einst\ Imaging Proportional Counter (IPC) data, and 
in LMC\,4 by Bomans, Dennerl, \& K\"urster (1994) with a 46 ksec 
\rosat\ Position Sensitive Proportional Counter (PSPC) observation.  
LMC\,2 is substantially brighter in X-rays than LMC\,4 and all other 
supergiant shells in the LMC. 

The physical and kinematic structure of LMC\,2 has been reported 
by Points et al. (1999; hereafter Paper~I).  It is found that two 
different mechanisms are responsible for generating the X-ray-emitting 
plasma: (1) local heating by supernova remnants (SNRs), e.g., the 
emission region enclosed by the northeast rim; and (2) outflows 
from active star formation regions, e.g., the bright X-ray arc 
extending from N\,158 (notation from Henize 1956) to N\,159 (or 
SNR\,0540-693 to LMC\,X-1; see Figure~1\,b).  The south end of LMC\,2 
is confusing.  Diffuse X-ray emission, while peaking within the 
shell, extends beyond the southern border of LMC\,2.  It has been 
speculated that this X-ray spur represents a blowout of LMC\,2, but 
no optical or radio supporting evidence has been found.

LMC\,2 clearly demonstrates the existence of hot gas interior to
a supergiant shell, and illustrates different mechanisms of generating
the hot gas.  Spatially resolved spectral analysis of the X-ray emission
within LMC\,2 and neighboring regions is the most effective way to 
determine the physical conditions of the hot plasma and establish its
role in the multi-phase interstellar medium.  

The X-ray emission from the interior of \lmc\ has not been studied 
since its initial detection with the \einst\ IPC.  Those IPC data were 
not used to make a detailed spectral analysis, however, because they 
had low spectral resolution and were contaminated by hard X-rays 
scattered from LMC~X-1 to large off-axis angles (Wang \& Helfand 1991).  
To study the distribution of X-ray emission toward \lmc, we 
have obtained \rosat\ PSPC and High Resolution Imager (HRI) observations
of \lmc.  Furthermore, to determine the physical conditions of the
X-ray emitting plasma interior to \lmc, we have also obtained \asca\ 
Solid-State Imaging Spectrograph (SIS) observations.

This paper is organized as follows: Section 2 describes 
the observations used in this study and their reduction.  
The distribution of the hot ($\ge 10^6$~K) plasma and its 
relationship to the warm ($\sim 10^4$~K) and cold ($\sim 
10^2$~K) interstellar component of \lmc\ is discussed in 
\S 3.  In \S 4 we describe the physical condition of the 
hot gas.  In \S 5 we compare the physical properties of
\lmc\ with other objects that exhibit diffuse X-ray emission.
We summarize our results in \S 6.

\section{Observations and Data Reduction}

The X-ray data reported in this paper were obtained using the \rosat\
PSPC and HRI and the \asca\ SIS.  

\subsection{X-ray Images}

The X-ray images of \lmc\ were obtained with the \rosat\ X-ray 
telescope, using both the PSPC and the HRI.  As discussed in 
Paper~I, we have mosaicked a large number of independent PSPC 
and HRI observations to provide extensive angular coverage, 
and considerably deeper exposure than provided by the \rosat\ 
All-Sky Survey (RASS).  The data were retrieved through the 
\rosat\ public archive at the High Energy Astrophysics Science 
Archive Research Center (HEASARC) at the Goddard Space Flight 
Center.  The region surrounding \lmc\ has been extensively 
covered by PSPC observations of the many sources in the vicinity 
(e.g., SNRs \& X-ray binaries) and by the HRI survey program of 
Chu \& Snowden (1998) which was designed to cover the 30~Doradus 
and bar regions of the LMC with deep and uniform coverage. Tables 
A1 and A2 list the observation sequences that were used for the 
mosaics of the PSPC and HRI, respectively.  These tables include 
the number of data sets for each sequence (individual observations 
were occasionally processed in sections), pointing direction, 
exposure time, and target name.  The data reduction was accomplished 
using the Extended Source Analysis Software (ESAS) package (Snowden 
\& Kuntz 1998; Kuntz \& Snowden 1998), which is also available 
through the HEASARC.  The final PSPC and HRI mosaics are cast in 
the same projection with the field center at $(\alpha,\delta)_{2000} 
= $ (5\rah 42\ram,$ -69\arcdeg 30'$) and pixel size of 30$''$ and 
20$''$, respectively.  The projection is the zenith equal-area 
azimuthal (ZEA; \cite{gre96}), which is similar over this solid 
angle to the more common tangential projection.

\subsubsection{PSPC Mosaic}
 
Reduction of the PSPC data followed the procedures outlined by Snowden 
et al. (1994) and demonstrated by Snowden \& Petre (1994), and in fact 
included a reprocessing of data in the latter paper.  Individual observations 
were screened for anomalous background conditions, had their residual 
non-cosmic background components modeled and subtracted, and had their 
vignetted and deadtime-corrected exposures calculated.  Only after all 
of the individual observations were reduced in six statistically-independent 
energy bands (see Table~1) were they cast into the mosaics.  Because the 
analysis software can not correct for zero-level offsets (a constant 
component of the long-term enhancement non-cosmic background, see 
\cite{smbm94}), the mosaicking process must correct for the relative 
offsets between overlapping fields.  A single-value deconvolution algorithm 
was applied to a system of equations comprising all overlaps between 
separate observations to determine a best fit for all offsets 
simultaneously.  A final constraint was added so that the sum of the 
offsets were equal to zero.  Individual observations were then adjusted 
by multiplying the fitted offset by the exposure map to produce an offset 
count image, which was then subtracted from the count image.  This 
procedure was run separately on each energy band.  Table~2 lists some 
values of interest for the reduction.  The true source counts can be 
seen to completely dominate the non-cosmic background and the average 
exposures are larger by more than an order of magnitude than that of 
the RASS coverage.  Although these source counts include emission 
from 30~Doradus, LMC~X-1, and various SNRs, \lmc\ is well above non-cosmic
background.  The total coverage of the field is $\sim18.2$~deg$^2$.  
We have extracted a $\sim 4.5$~deg$^2$ sub-image of the \lmc\ region 
from the larger mosaic and present it in Figure~1\,a.

\placetable{tbl:pspc-bands}

\placetable{tbl:stats}
 
\placefigure{fig:fig1ab}

\subsubsection{HRI Mosaic}
 
The reduction of HRI data is simplified from that of PSPC data.  
There is only the total energy band, and only the particle background 
is modeled and removed.  The particle background of the HRI completely 
dominates an HRI observation with an average count rate $\sim5-10$ 
times greater than the cosmic background.  It must be removed as the 
distribution of the particle background is flat across the field while 
cosmic counts are vignetted.  Unfortunately, the particle background 
is also only calibrated to $\sim10$\% or so, but this is sufficient 
for the purposes of this paper.  The contributions from other non-cosmic 
background components are ignored as they are ``in the noise.''  Because
these components are for the most part distributed as a vignetted flat 
field, they contribute a zero-level offset which is corrected in the 
mosaicking, which is done in the same manner as for the PSPC.  The total 
coverage of the field is $\sim8.1$~deg$^2$.  In Figure~1\,b we present a 
sub-image from this mosaic that has the same field-of-view as Figure~1\,a.

\subsection{X-ray Spectra}

The X-ray spectra in this paper were obtained with the \rosat\ 
PSPC and the \asca\ SIS.  Here we discuss the reduction of the 
X-ray spectral data.  We discuss their analysis in \S 4.

\subsubsection{\rosat\ PSPC Spectra}

We extract X-ray spectra of \lmc\ from a \rosat\ PSPC observation
of the SNR DEM~L~316 (\rosat\ Sequence No. RP500259N00). The PSPC 
is sensitive in the energy range of 0.1 -- 2.4~keV, and has an 
energy resolution of $\approx 43$\% at 1~keV.  This observation 
has been discussed in some detail in Williams et al. (1997).  
Therefore, we only present a brief summary here.  The observation 
was made within the time interval from 1993 January 10 to 1993 
February 10.  For this observation, the SNR DEM~L~316 ($\alpha_{2000} 
= $ 5\rah 47\ram 9\fs 5; $\delta_{2000} = -69\arcdeg 42\arcmin 
14\arcsec$) was centered in the PSPC.  The total exposure time for 
the observation was 4.0~ksec.

We processed these data using Xselect and FTOOLS\footnote{These data 
were also reduced using the IRAF/PROS packages.  No apparent differences 
exist between the resulting spectra.}, which are available through 
the HEASARC.  The data were spatially binned to produce 7\farcs 5 
$\times$ 7\farcs 5 pixels.  Using the binned PSPC image, we defined 
three regions of interest: a bright region in the northern part of 
\lmc\ (PSPC-North), a bright X-ray arc in the west (PSPC-Arc), and a 
region in the X-ray spur to the south of \lmc\ (PSPC-Spur).  These regions 
are shown in Figure~1\,a.  Because \lmc\ is an extended source and 
all the regions are located outside the central PSPC ``ring'' (radius 
20$'$), we selected background regions of the same size and off-axis 
angle to correct for the charged particle background.  During the 
reduction process we also performed vignetting corrections.  We
present the regions' coordinates, angular sizes, net counts, and
effective exposure times in Table~3.

\placetable{tbl:reg-prop}

\subsubsection{\asca\ SIS Spectra}

Our \asca\ observation is of a bright X-ray region in the northern
section of \lmc\ (see Figure~1\,a). This observation was performed on 
1995 August 25 -- 26  (\asca\ Sequence No. 53039000).  

The \asca\ satellite consists of two instruments that observe 
simultaneously -- a pair of SIS and a pair of Gas Imaging 
Spectrographs (GIS).  The SIS has an energy coverage approximately 
in the 0.5 -- 8~keV range and the GIS in the 0.8 -- 10~keV range 
(Tanaka, Inoue, \& Holt 1994).  In this paper, we have only used 
the SIS data because of the low sensitivity of the GIS for energies 
below $\approx 1.0$~keV, where diffuse thermal plasma emission is 
brightest.  The SIS detectors have good spectral resolution 
$[\delta E/E \sim 0.02(5.9~{\rm keV}/E)^{0.5}]$.  The angular 
resolution of the SIS data is poor compared to the \rosat\ PSPC, 
with a narrow core of $\approx 1'$ diameter and a half-power radius 
of 3$'$.  Both of the SIS detectors (SIS~0 and SIS~1) are composed 
of four separate CCDs.  Not all of the CCDs can be on for a given 
observation as the presence of ``hot'' and ``flickering'' pixels 
can fill the telemetry with false signals.

Because of telemetry limitations, our observations were performed 
in the 2-CCD mode, with part of the bright northern region of \lmc\ 
in the SIS~0 (Chips 0 \& 1) and SIS~1 (Chips 0 \& 3).  We have 
extracted spectra from SIS~0 Chip~1 and SIS~1 Chip~3, because they 
image the same area on the sky and lie in a region of bright emission.  
We reduced and processed the data with ASCASCREEN and FTOOLS.  
Because \lmc\ is more extended than the SIS field-of-view no region
from any of our \asca\ observations were suitable to perform a
background subtraction.  Therefore, we used the \asca\ Deep Field 
pointings, screened in a manner identical to the \lmc\ data, to obtain 
regions for background subtraction. In Table~3, we also present relevant 
information on the two \asca\ SIS Chips that were used in this paper.
We have also extracted spectra in the regions covered by SIS~0 Chip~0 
and SIS~1 Chip~0; however, these data were not analyzed because 
the count rates were low and these spectra cannot be fitted by models
with reasonable certainty.

\section{Spatial Distribution of the ISM toward \lmc}

\lmc\ consists of interstellar gas in several phases.  As reported 
in Paper~I, we have obtained (1) \rosat\ PSPC and HRI mosaics to 
investigate the spatial distribution of the hot ionized gas interior 
to \lmc, (2) CCD images of \lmc\ in optical emission lines to examine 
the shell of warm ionized gas, and (3) 21-cm \hi\ data to study the
structure of the cold, neutral gas.  Below we describe the distribution 
of the X-ray emitting plasma and its relation to the other gaseous 
components.

\subsection{Morphology of the Hot Ionized Medium of \lmc}

The existence of hot ($\ge 10^6$~K) gas interior to \lmc\ was 
first detected using the \einst\ IPC (Wang \& Helfand 1991).
As seen in our \rosat\ PSPC and HRI mosaics (Figures~1\,a \& 1\,b, 
respectively), \lmc\ lies toward the brightest and most complex
region of diffuse X-ray emission in the LMC.  This region includes 
30~Doradus, \lmc, the supergiant shell LMC\,3, and a bright X-ray 
spur to the south of \lmc, as well as SNRs (e.g., DEM~L~299, 
DEM~L~316, and SNR 0540$-$69.3) and an X-ray binary (LMC~X-1).  
These features are labeled on our HRI mosaic in Figure~1\,b.

A cursory inspection of our PSPC and HRI mosaics reveals that the 
X-ray surface brightness of \lmc\ is not uniform.  The brightest 
region of diffuse X-ray emission associated with \lmc\ is an arc 
in the southwestern quadrant that appears to extend from SNR~0540$-$693 
to LMC~X-1.  A region of low X-ray surface brightness ($\alpha_{2000} 
= 5^{\rm h} 45^{\rm m}$, $\delta_{2000} = -69^{\circ} 25'$) lies 
between this bright X-ray arc and a region of bright X-ray emission 
in the northeast.  The observed variation in X-ray surface brightness 
can be attributed to non-uniform emission from \lmc\ and/or to 
non-uniform absorption of X-rays by intervening interstellar gas.  
As discussed below, we find that both of these explanations must be 
taken into account for the observed X-ray surface brightness toward 
\lmc.

\subsection{Comparison of X-ray Images with Optical/Radio Observations}

It is of considerable importance in our study of the physical 
conditions interior of \lmc\ to determine whether the variation 
in X-ray surface brightness is the result of non-uniform 
absorption or emission.  To investigate this matter we first 
compare the X-ray mosaics with \ha\ emission line images that
trace regions of current massive star formation.  As the hot 
gas toward \lmc\ is most likely shock-heated by the stellar 
winds and supernova explosions of massive stars, we expect 
to see correspondence between X-ray and \ha\ features.  We 
present an \ha\ image of \lmc\ overlaid with X-ray contours 
in Figure~2.

\placefigure{fig:fig2ab}  

We find three results of note in this comparison: (1) the X-ray 
emission from \lmc\ is confined by the optical filaments; (2) 
the bright X-ray arc is centered on the \hii\ region N\,160.  
Based on the optical and X-ray morphology and high resolution 
echelle spectra of the \ha\ line, we suggest that the arc is 
formed by an outflow of hot gas from N\,160 into \lmc\ (Paper~I).  
(3) The \hii\ region N\,163 ($\alpha_{2000} = 5^{\rm h} 43^{\rm m}
09^{\rm s}, \delta_{2000} = -69\arcdeg 45\arcmin 54\arcsec$) is 
coincident with a small region of low X-ray surface brightness.  
Comparison of the PSPC mosaics (Snowden \& Petre 1994) shows 
that this depression in X-ray surface brightness is more 
pronounced in the soft X-ray band (R4) than in the hard X-ray 
band (R7).  Therefore, we conclude that N\,163 lies in front 
of the X-ray emitting gas and absorbs the soft X-rays from \lmc.

We further compare the X-ray mosaic with the Australia Telescope 
Compact Array (ATCA) aperture synthesis maps of the 21-cm \hi\ 
emission line toward \lmc\ that reveals the presence of neutral 
atomic gas (Figure~3).  We know that \hi\ exists in front and in 
back of \lmc.  Thus, we cannot simply expect variations in the 
\hi\ column density to anti-correlate with the diffuse X-ray 
emission.  Instead we use the shadowing effect to look for 
correlations between features in the \hi\ channel maps with X-ray 
surface brightness.  As reported in Paper~I, a region of low X-ray 
surface brightness ($\alpha_{2000} = 5^{\rm h} 45^{\rm m}, \delta_{2000} 
= -69\arcdeg 25\arcmin$) is coincident with an \hi\ cloud detected 
in the \hi\ channel maps.  As in the case of the region with low 
X-ray surface brightness seen toward N\,163, this depression is more 
apparent in the R4 band mosaic than the R7 band mosaic.  Therefore, 
this \hi\ gas component most likely lies on the front side of 
\lmc\ and absorbs the X-rays emitted from within \lmc\ (Paper~I).

\placefigure{fig:fig3}

\section{Physical Conditions of the Hot Gas}

The \rosat\ PSPC and \asca\ SIS data provide X-ray spectra of \lmc\
that can be used to derive the physical conditions of the hot gas.  We
will first describe the procedures used to perform the X-ray spectral
fitting and the uncertainties involved.  Then we will use the plasma
temperature, foreground absorption column density, and X-ray
emissivity obtained from the spectral fits to estimate the electron
density, surface mass density, and surface thermal energy density of
the hot gas.

\subsection{X-ray Spectral Fits}

The observed X-ray spectra are a convolution of the intrinsic 
spectra, the foreground absorption, and the detector response
function.  Because the foreground absorption and detector
response are both energy dependent, we need to assume models
of the intrinsic spectra and absorption cross-sections to 
reduce the number of free parameters.  We simulate the observed
spectra by folding the assumed intrinsic emission and foreground
absorption models with the detector response.  The observed 
spectra are fitted by the simulated spectra; the $\chi^2$ of 
the fits is used to determine the best-fit.

The X-ray emission from \lmc, as seen in the HRI mosaic
(Figure~1\,b), is truly diffuse, indicating a hot-plasma
origin.  Therefore, we adopt Raymond \& Smith's (1977) thin 
plasma emission models for the intrinsic spectra.  We use 
Morrison \& McCammon's (1983) effective absorption cross-sections 
per hydrogen atom to model the foreground absorption.  The 
PSPC and SIS data were fitted separately because they were
extracted from different regions toward \lmc.

The spectral fitting of the \asca\ SIS data was more difficult
than the fitting of the \rosat\ PSPC data.  The SIS data were
contaminated by scattered, high-energy photons from the bright
nearby source LMC~X-1, an X-ray binary with an accretion disk.  
Therefore, in the fitting procedure for the SIS data we added 
a powerlaw and a blackbody disk component, but held the values 
of the power law photon index and the blackbody disk temperature 
constant at the values determined by Schlegel et al. (1994) for 
LMC\,X-1.  We allowed the values of the Raymond-Smith plasma 
temperature and absorption column to float to the best-fit values.  
This is not a perfect method to determine the plasma temperature 
and absorption column of \lmc\ because the scattering of the 
X-ray photons is both energy- and position-dependent.  We may 
fit the SIS~0 Chip~1 and SIS~1 Chip~3 data simultaneously 
with the same model because these data were extracted from the 
same region on the sky.

We performed multiple model fits for each of the extracted 
regions using XSPEC in which we allowed the abundance to be 
20\%, 30\%, and 40\% solar.  Although varying the abundance 
did not noticeably affect the best-fit values of the plasma 
temperature and absorption column, it did change the parameter 
space of the $\chi^2$ grid plots of the confidence contours.  
We find that the range of possible values for the plasma 
temperature and absorption column are better constrained for 
the models with 40\% solar abundance.  For the remainder of 
this work, however, we use the results from the 30\% solar 
abundance models, the canonical LMC value.  Although our 
choice to use the best-fit values from the 30\% solar abundance 
models may seem arbitrary, this allows us to compare our results 
with other studies of the hot, ionized medium in the LMC (\S 5).  
We note that the $\chi^2$ grid plots for a thermal plasma with 
40\% solar abundance may cover a smaller area of parameter space 
because the hot gas has been enriched above the canonical LMC 
value through the actions of stellar winds and supernova 
explosions of massive stars.

We performed a $\chi^2$ grid search of possible fits in order 
to determine the uncertainties in the fits for temperature ($kT$) 
and absorption column density ($N_H$).  The best-fit plasma 
temperature and absorption column for a 30\% solar abundance 
Raymond \& Smith (1977) thermal plasma model for PSPC and SIS 
data, as well as the range of the 90\% confidence contours, are 
given in the top section of Table~4.  In Figures~4 \& 5, we present 
the X-ray spectra of the different regions, and their $\chi^2$ 
grid plots for a thermal plasma with 30\% solar abundance, respectively.  
The grid plots show the range of $\chi^2$ fits for $kT$ and $N_H$ 
for each extracted region; the three contour levels represent the 
areas of 68\%, 90\%, and 99\% confidence.  

\placetable{tbl:xray-fit}

\placefigure{fig:fig4}

\placefigure{fig:fig5}

To better constrain the range of $N_H$, and thus to better confine the
range of plasma temperatures, we calculate the \hi\ column density
toward \lmc.  The Galactic values of $N_H$ toward \lmc\ are $\sim 6.5
\times 10^{20}$~cm$^{-2}$, determined using the FTOOL ``nh'' which is
derived from the \hi\ map of Dickey \& Lockman (1990).  The LMC values
of $N_H$ toward \lmc\ are $\sim 4 \times 10^{21}$~cm$^{-2}$ using the
Parkes Multibeam \hi\ Survey.  The best-fit values of $kT$ for the
fixed $N_H$ model fits, as well as the 90\% confidence ranges for $kT$
and $N_H$ are given in the bottom section of Table~4.  In general, the
fixed $N_H$ column density is lower than that determined by the
floating $N_H$ X-ray spectral fits (see Table~4), similar to results
obtained toward 30~Doradus (Wang 1999).  This appears to be
problematic because the Parkes Multibeam Survey samples neutral \hi\
that is both in front of and behind \lmc.  It seems reasonable that
the fixed $N_H$ column density should represent an upper limit of the
absorption column between us and the X-ray emitting plasma.  It should
be remembered, however, that X-rays are also absorbed by molecular and
ionized gas that are not measured in the 21-cm \hi\ emission-line
survey, but are measured by the X-ray spectral fits.  Observations of
the $J = 1-0$ CO emission line at 2.6~mm made with the NANTEN 4-m
radio telescope (Fukui et al. 1999) do not show CO emission toward the
\lmc\ regions of interest in this investigation.  The lack of CO
emission does not necessarily imply an absence of molecular gas toward
the selected regions.  The lower dust abundance of the LMC allows UV
photons, which would dissociate CO molecules, to penetrate more deeply
into molecular clouds and form C$^+$ regions.  Observations of the
[C\,{\sc ii}] 158$\mu$m emission line toward \lmc\ reveal that the
distribution of C$^+$ is more diffuse than the distribution of CO
(Mochizuki et al. 1994).  Therefore, it is possible that the molecular
gas is in a very diffuse phase in which CO is rare because of
photo-dissociation.  Our calibrated CCD \ha\ images of \lmc\ (Paper~1)
show the emission measure, $N_e^2 L$, where $N_e^2$ is the electron
density and $L$ is the path length through the \ha\ emitting gas in
pc, range from $\sim 250$~cm$^{-6}$~pc in the arc region to $\sim
20$~cm$^{-6}$~pc in the spur region.  If the path length through the
\ha\ emitting gas is comparable to the width of the filaments defining
the eastern boundary of \lmc\ ($\sim 5$~pc), the ionized column
densities toward the regions range from $N_{H^+} = 3.1 \times
10^{19}$~cm$^{-2}$ in the spur to $N_{H^+} = 1.1 \times
10^{20}$~cm$^{-2}$ in the arc.  It is conceivable that the derived
absorption column from the free $N_H$ X-ray spectral fits may be
unrealistically higher than the measured absorption column toward the
X-ray spectral regions.  Therefore, we use the fixed $N_H$ spectral
fit parameters for the remainder of this work.

\subsection{X-ray Surface Brightness and Luminosity of \lmc}

From the model fits to the PSPC and SIS X-ray spectra, we can 
calculate the unabsorbed X-ray flux, and hence the X-ray 
luminosities, of the selected regions.  Given a hot gas filling 
factor $f$, volume $V$, X-ray luminosity $L_X$, and emissivity 
$\Lambda$, the electron density of the hot plasma of the selected 
regions will be
$$n_e = (1.1 L_x)^{1/2} (\Lambda V f)^{-1/2},$$
if $n_e = 1.1 n_H$.  For the regions toward \lmc\ and the spur, the
volume of the gas is the product of the projected surface area of the
region and the path length through the gas.  We adopt a path length
comparable to the width of the interstellar structure in question
(i.e., 140~pc for the bright northern region, 230~pc for the bright
X-ray arc, and 400~pc for the spur).  The X-ray emissivity, $\Lambda$,
is a function of the plasma temperature.  From our fixed $N_H$ X-ray
model fits to the data, we find that the emissivity of the regions
ranges from $5 - 7 \times 10^{-24}$~erg~cm$^{3}$~s$^{-1}$ in the 0.44
-- 2.04~keV energy range.  We derive the electron densities for the
extracted regions toward \lmc\ using the X-ray luminosities and
emissivities determined by our fixed $N_H$ spectral fits and present
them in Table~5.

\placetable{tbl:xray-prop}

As seen in Figure~1\,a, the X-ray surface brightness varies from
region to region toward \lmc.  Because \lmc\ covers a large portion of
the PSPC field-of-view, parts of it are obscured by the PSPC window
support structure.  Thus, we cannot extract a single X-ray spectrum of
\lmc\ to determine the integrated X-ray flux of the hot gas.  Instead,
we determine the total count rate of \lmc\ from the PSPC mosaic in the
0.44 to 2.04~keV energy band.  If we assume that \lmc\ is filled by
hot gas with a plasma temperature $kT = 0.31$~keV and an absorption
column of $N_H = 4.5 \times 10^{21}$~cm$^{-2}$, we can use
PIMMS\footnote{PIMMS is made available from the HEASARC at {\it
http://heasarc.gsfc.nasa.gov/docs/software/tools/pimms.html}.}  to
calculate the intrinsic X-ray flux of \lmc\ to be $1.4 \times
10^{-10}$~erg~cm$^{-2}$~s$^{-1}$.  The intrinsic X-ray flux of the
spur is calculated to be $1.0 \times 10^{-10}$~erg~cm$^{-2}$~s$^{-1}$
for a thermal plasma with temperature $kT = 0.22$~keV and a hydrogen
absorption column of $N_H = 4.5 \times 10^{21}$~cm$^{-2}$.  We caution
the reader, however, that these derived values of the X-ray flux were
made assuming a uniform absorption column between us and \lmc\ and the
spur.  Therefore, these values do not accurately reflect the global
X-ray fluxes of either of these interstellar structures.

\subsection{Derived Physical Properties}

As X-ray fluxes have now been calculated for both \lmc\ 
and the spur, we may be tempted to derive their physical 
conditions, such as electron density ($n_e$), thermal 
energy ($E_{th}$), and hot gas mass ($M$).  This approach, 
however, would ignore the uncertainties inherent in the 
assumptions we made in our previous calculation of the 
intrinsic X-ray flux of \lmc\ and the spur, such as 
the emitting volume, temperature, and the emissivity of 
the hot, thermal plasma.  Furthermore, we have shown in 
\S3 that the absorption column across \lmc\ is non-uniform.
Therefore, the  physical conditions of the hot gas interior
to \lmc\ and the spur cannot be realistically derived 
using the global X-ray fluxes calculated in \S4.2.  Instead, 
we opt to focus on the physical conditions of the hot gas
toward the extracted spectral regions because the plasma 
temperatures, X-ray fluxes, and hot gas emissivities are 
explicitly determined by the X-ray spectral fits. To further 
minimize uncertainties in the derived physical conditions of 
the hot gas toward our extracted spectral regions, we 
calculate the surface thermal energy density ($\Sigma_{E_{th}}$)
and the hot gas surface mass density ($\Sigma_M$) instead
of $E_{th}$ and $M$ because they depend on the assumed path 
length through the hot gas, but not the volume.

In the following analysis we derive the physical properties of the hot
gas in our selected regions toward \lmc.  As an example, we calculate
the physical properties of the hot gas in the PSPC-Arc region.  The
electron density of this region is $n_e = (0.06 - 0.08)$~cm$^{-3}$.
The surface mass density of the hot gas is given by $\Sigma_M = 1.4
m_H n_H V f / \Omega$ where $Vf$ is the effective volume of the
emitting gas and $\Omega$ is the surface area of the region in
arcmin$^{2}$.  Assuming that $n_e = 1.1 n_{H}$, the surface mass
density of the hot gas in the arc ranges from 65 to
90~\Msun~arcmin$^{-2}$.  If the X-ray emitting gas behaves as an ideal
gas, the surface thermal energy density is given by $\Sigma{E_{th}} =
{3 \over 2} n_e V f k T/ \Omega $.  For the arc, this gives
$\Sigma_{E_{th}} = (0.5 - 0.7) \times 10^{50}$~erg~arcmin$^{-2}$.
Furthermore, we may also estimate the number of supernova explosions
required to produce the hot gas interior to \lmc, based on our derived
value of $\Sigma_{E_{th}}$.  For a supernova explosion with a total
energy of $5.0 \times 10^{50}$~erg, the thermal energy of the hot gas
of the SNR is $0.2 \times 10^{50}$~erg and has a radius of $\sim
60$~pc after 10$^6$~yr (Slavin \& Cox 1992).  At the distance to the
LMC, the surface thermal energy density of the SNR would be
$\Sigma_{E_{th}} \approx 0.004 \times 10^{50}$~erg~arcmin$^{-2}$.
Thus, if the hot gas interior to \lmc\ was produced entirely by a
series of supernova explosions $\approx 10^6$~yr ago, then the number
of supernova explosions required to energize the arc is 125 -- 175.
We have obtained UBV photometry to investigate the star formation
history toward \lmc.  Those data will allow us to determine the amount
of energy massive stars have deposited in the ISM of \lmc\ and will be
compared with the thermal energy that is derived here.  The physical
properties of the arc, as well as those for the other regions toward
\lmc, are presented in Table~6.

\placetable{tbl:xray-para}

Before comparing the physical properties of the hot gas interior to
\lmc\ with the physical properties of hot gas in other interstellar
structures, we re-iterate the assumptions that underlie our
calculations.  The first factor we consider that contributes to the
uncertainty in our calculations is the volume of the X-ray emitting
gas in each of the regions.  In this work we simply assume that the
volume of a region is the product of its projected surface area and
the path length through the region.  Because the depth through these
regions is uncertain, an increase of the path length by a factor of 2
would alter our determination of the electron density by factor of
$1\over{\sqrt 2}$ and our calculations of the the surface thermal 
energy density and mass by a factor of $\sqrt 2$.  Another factor
affecting the derivation of the physical properties of the hot gas
interior to \lmc\ is the volume filling factor, $f$.  We have
determined the physical conditions of the hot gas using a filling
factor of $f = 0.5$ and $1.0$.  A filling factor of $f = 0.5$
increases the derived values of n$_e$ by a factor of $\sqrt 2$ and
decreases the values $\Sigma_{E_{th}}$ and $\Sigma_M$ by a factor of
$1\over{\sqrt 2}$, with respect to the values determined using a
filling factor of $f = 1.0$.  Inspection of Figures~1\,a \& 1\,b
reveals that the distribution of the X-ray emission is relatively
smooth and does not show much structure.  Thus, it is likely that the
value of the filling factor of the hot gas interior to \lmc\ lies
between these two bounds.  The last factors that need to be considered
as contributing to the uncertainty in our results are the best-fits to
the plasma temperature, $kT$, the absorption column, $N_H$, and the
normalization constant, $C$, defined as $C = {{10^{-14} \int n_e n_H
dV}\over{4 \pi d^2}}$.  As seen in the $\chi^2$ grid plots (Figure~5),
the confidence contours for the PSPC data show that the thermal plasma
spectrum can be fitted by a low plasma temperature and high absorption
column or by a high plasma temperature and a low absorption column.
For example, if we take the plasma temperature of the PSPC-North
region to be the free $N_H$ model best-fit ($kT = 0.15$~keV; $N_H =
8.6 \times 10^{21}$~cm$^{-2}$), the derived values of $n_e$ and
$\Sigma_M$ are $\sim 7.5$ times the values derived using the fixed
$N_H$ model fit values ($kT = 0.31$~keV; $N_H = 4.8 \times
10^{21}$~cm$^{-2}$).  In this example, the surface thermal energy 
density is $\sim 3.5$ times the value calculated for the fixed $N_H$
model fits and the intrinsic X-ray flux, $F_X$ is increased by a
factor of $\sim 20$.

\section{Physical Properties of Other Objects with Diffuse
X-ray Emission}

The hot gas in the ISM is produced by shocks from supernova
blasts and fast stellar winds from massive stars.  Consequently,
the amount of hot gas and its physical conditions depend upon
the spatial distribution and formation history of massive stars.
Massive stars will form discrete SNRs if they are isolated, superbubbles
if there is a single burst of star formation such as an OB association
or supergiant shells if there are multiple bursts of star formation 
that have taken place over a period of time in a concentrated 
area.  SNRs, superbubbles, and supergiant shells are all filled 
by hot gas which contributes to the HIM (hot ionized medium; McKee \&
Ostriker 1977) component of the ISM in galaxies.  Thus, it is of 
interest to compare the physical conditions of the hot gas in these 
interstellar structures.

Below we compare the physical properties of the hot gas interior 
to \lmc\ with those of: (1) another supergiant shell in the LMC,
LMC\,4; (2) two superbubbles in the LMC, N\,11 and N\,44; and (3)
a number of SNRs in the LMC.  We present the physical properties 
of these objects in Table~7.  The values of the physical properties 
for these objects were made using the \rosat\ bandpass of 0.1 -- 
2.4~keV and assuming a hot gas filling factor of $f = 1.0$ unless 
otherwise noted.

\placetable{tbl:xray-other}

\subsection{Supergiant Shells in the LMC}

The pioneering study of diffuse X-ray emission from within a
supergiant shell in the LMC was performed on \lmc\ by Wang \& Helfand
(1991) using data obtained with the \einst\ IPC which operated in the
energy range from 0.4 to 4.0~keV.  Therefore, as a consistency check,
we first compare our results of the physical conditions of the hot gas
interior to \lmc\ to those which they derive.  Because a detailed
spectral analysis of \lmc\ could not be performed using the IPC data,
Wang \& Helfand (1991) used the mean hardness ratio of the excess
X-ray emission from \lmc\ to estimate the temperature of the hot
gas. By assuming an optically thin thermal plasma with 50\% solar
abundance for the emission and a foreground absorption of $N_{H} \sim
3 \times 10^{21}$~cm$^{-2}$, they determine the temperature of the
X-ray emitting gas to be $\sim 0.43$~keV.  Wang \& Helfand (1991)
assume the geometry of \lmc\ to be spherical and adopt two different
radii to determine the volume and physical properties of the \him\
gas: (1) a radius of $250$~pc that covers only the arc region, and (2)
a radius of $500$~pc that encompasses the entire supergiant shell.
They find the density of the hot gas to be $n_e \sim 2 \times
10^{-2}$~cm$^{-3}$ or $n_e \sim 9 \times 10^{-3}$~cm$^{-3}$ for an
adopted radius of 250~pc or 500~pc.  Their derived values of the
unabsorbed X-ray flux, $F_X$ and the electron density, $n_e$ are given
in Table~7.  Using their derived values, we calculate the surface 
thermal energy density, $\Sigma_{E_{th}}$, to be $0.46 \times
10^{50}$~erg~arcmin$^{-2}$ and $0.41 \times 10^{50}$~erg~arcmin$^{-2}$
for a spherical bubble of hot gas with a radius of 250~pc and 500~pc,
respectively.

The results obtained using the \einst\ IPC data are in good 
agreement with those we determined using the \rosat\ PSPC data, 
but not with the SIS data.  For example, their values for the 
plasma temperature ($kT = 0.43$~keV) and the absorption column 
($N_H = 3 \times 10^{21}$~cm$^{-2}$) fall within the 90\% 
confidence contour limits of our $\chi^2$ grid-plots of the 
PSPC data, but not the SIS data (see Figure~5).  A factor that 
may produce a discrepancy between the X-ray spectral fits to 
our dataset and the analysis of the IPC data is that Wang \& 
Helfand (1991) assume a uniform hydrogen column density toward 
\lmc.  We have shown in \S3.2 that the variations in X-ray 
surface brightness toward \lmc\ are caused by non-uniform 
absorption and emission.  As discussed previously, a higher
adopted absorption column will result in a lower value for
the plasma temperature.  Therefore, it is likely that the 
differences between the observed spectral properties and the
derived physical conditions can be attributed to the differences
in the adopted absorption column. 

In addition to \lmc, diffuse X-ray emission has also been 
detected in the northern quadrant of the supergiant shell 
LMC\,4 (Bomans, Dennerl, \& K\"urster 1994) using the \rosat\ 
PSPC in the energy range between 0.1 and 2.4~keV.  They find 
that the best fit Raymond \& Smith (1977) plasma temperature 
to the data is $kT = 0.21$~keV and the best fit absorption 
column is $N_H = 0.7 \times 10^{21}$~cm$^{-2}$.  If the 
diffuse X-ray emission from the northern quadrant of LMC\,4 
is representative of LMC\,4 as a whole, the X-ray flux is 
$\sim 5 \times 10^{-11}$~erg~cm$^{-2}$~s$^{-1}$ for this 
supergiant shell.  Assuming that LMC\,4 is a cylinder with 
a radius of 600~pc, a height of 1200~pc, and an X-ray emissivity
of $\Lambda = 4.0 \times 10^{-24}$~erg~cm$^{3}$~s$^{-1}$, 
they derive an electron density of $n_e = 8 \times 10^{-3}$~cm$^{-3}$.  
Using these values we calculate the surface thermal energy 
density of LMC\,4 to be $\Sigma_{E_{th}} = 0.32 \times 
10^{50}$~erg~arcmin$^{-2}$.  The derived values of $F_x$, 
$n_e$, and $\Sigma_{E_{th}}$ are listed in Table~7.

The plasma temperature of the hot gas interior to LMC\,4 
($kT = 0.21$~keV) is comparable to that we have determined 
for \lmc\ at the 90\% confidence level, but the absorption 
column is about an order of magnitude lower.  The primary 
distinction between the X-ray emitting plasma in \lmc\ and 
LMC\,4 is that the electron densities in the regions toward 
\lmc\ are a factor of $\sim 10$ greater than that derived for 
LMC\,4.  As we have used the same filling factor and a similar 
X-ray emissivity for the $\ge 10^6$~K gas as Bomans et al. 
(1994), the difference in the derived electron densities can 
be ascribed to the higher unabsorbed X-ray flux determined 
for \lmc\ and/or through uncertainties in the volume of the 
emitting thermal plasma.  If we adopt a path length through 
our selected regions toward \lmc\ comparable to that of LMC\,4 
(i.e., $l= 1200$~pc), our derived electron densities would be 
lowered only by a factor of $\sim 3 - 5$.  Therefore, we conclude
that the electron density of the hot gas interior to \lmc\ is 
generally higher than the electron density interior to LMC\,4.

\subsection{Superbubbles in the LMC}

Diffuse X-ray emission has been detected in several LMC 
superbubbles.  Here we discuss the physical conditions 
of the hot gas interior to the superbubbles in N\,11 and
N\,44, the two most luminous \hii\ regions in the LMC 
after 30 Doradus (Kennicutt \& Hodge 1986).  N\,44 has 
the highest X-ray surface brightness among all superbubbles
in the LMC (Chu \& Mac Low 1990), while N\,11 is only 
moderately X-ray bright (Mac Low et al. 1998).  The 
physical properties of the hot gas interior to these two 
superbubbles are given in Table~7.  

Raymond \& Smith (1977) spectral model fits to \rosat\ PSPC 
observations of N\,11 give a best-fit plasma temperature of 
$kT = 0.3$~keV (Mac Low et al. 1998).  This plasma temperature 
is higher than the model spectral fits to our PSPC data toward 
\lmc, but is comparable to plasma temperatures determined by 
the model fits to the SIS data.  Using their results we calculate 
the electron density of N\,11 to be $n_e = 0.06$~cm$^{-3}$ 
and the surface thermal energy density to be $\Sigma_{E_{th}} = 
0.23 \times 10^{50}$~erg~arcmin$^{-2}$.  

Both PSPC and SIS data have been obtained for N\,44 (Magnier 
et al. 1996).  Because these observations cover the same region 
on the sky, the PSPC and SIS X-ray spectra were fit simultaneously 
with a thermal plasma emission model.  The model fits give a 
plasma temperature of $kT = 0.55$~keV for the hot gas interior 
to N\,44 (Magnier et al. 1996).  Assuming a filling factor of 
$f = 0.5$, they derive an electron density of $n_e = 0.14$~cm$^{-3}$ 
from which we calculate the surface thermal energy density to 
be $\Sigma_{E_{th}} = 0.07 \times 10^{50}$~erg~arcmin$^{-2}$.  

The plasma temperatures of the hot gas interior to N\,11 is similar 
to those we determine for the hot gas interior to \lmc\ using the 
PSPC and SIS data, but are higher in N\,44. The electron density 
and surface thermal energy density for N\,11 are on the same order 
as for the regions toward \lmc.  The electron density of the hot 
gas interior to N\,44 is a factor of $\sim 1.5 - 3$ times higher 
than that derived for \lmc, but the surface thermal energy density 
is $\sim 4 - 11$ times lower.  Some of the difference between the 
derived physical conditions of the hot gas interior to \lmc\ and 
N\,44 may be attributed to the fact that Magnier et al. (1996) used 
a hot gas filling factor of $f = 0.5$.  The effects of changing the
filling factor from $f = 1.0$ to $f = 0.5$ are discussed in \S 4.3. 

\subsection{SNRs in the LMC}

Supernova explosions are the main producers of the HIM in 
galaxies.  Therefore, it is imperative to understand the
physical properties of the hot gas interior to SNRs, in 
order to understand the physical conditions of hot gas 
interior to superbubbles and supergiant shells.  Fortunately, 
the LMC contains a large number of SNRs ($\sim 40$; Williams 
1999) that have been studied extensively at optical and radio 
wavelengths (Mathewson et al. 1983; 1984; 1985) and X-ray 
wavelengths (Williams et al. 1997; 1999; Williams 1999; 
Hughes et al. 1998).  Williams (1999) has compiled the most 
comprehensive list of the X-ray properties of LMC SNRs to 
date.  Of the LMC SNR sample, twenty-three have \rosat\ PSPC 
data available with sufficient count rates for spectral 
modeling.  For this dataset, Williams (1999) finds that the 
plasma temperature ranges from $kT  = 0.14$ to 0.89~keV 
and that the electron density ranges from $n_e = 0.05$ to 
$\sim 20$~cm$^{-3}$.  The published physical conditions of 4 
of these SNRs are listed in Table~7.

Clearly, the hot gas interior to SNRs in the LMC shows a 
wide range of physical conditions.  This is not surprising 
as they have different ages and interstellar environments.
We see that most of these LMC SNRs have plasma temperatures 
that are higher than \lmc, but have electron densities that 
are comparable to it.

\section{Summary}

The supergiant shell \lmc\ has the highest X-ray surface 
brightness of all LMC supergiant shells.  We have obtained
\rosat\ PSPC and HRI mosaics to study the distribution of 
X-ray emission from \lmc.  Comparison of the X-ray mosaics 
with \ha\ images show that the X-ray emission is detected 
within the shell boundary of \lmc, indicating the presence 
of hot ($\ge 10^6$~K) gas.  In general, regions of low X-ray
surface brightness are well-correlated with regions of high
\hi\ column density, suggesting that some of the variations
in X-ray surface brightness toward \lmc\ are caused by non-uniform
absorption.

In addition to the X-ray mosaics, we have also obtained \rosat\ 
PSPC and \asca\ SIS X-ray spectra to determine the physical
conditions of the hot gas interior to \lmc.   Model fits of an 
optically thin thermal plasma (Raymond \& Smith 1977) to the 
X-ray spectra of selected regions show that the hot gas has a 
plasma temperature $kT = 0.1$ -- $0.4$~keV with a best-fit of 
$kT \sim 0.3$~keV for our PSPC and SIS data.  The electron density 
of the hot gas varies from $n_e \sim 0.05$ to 0.09~cm$^{-3}$.  The 
unabsorbed X-ray flux in the 0.44 -- 2.04~keV band is $1.4 \times 
10^{-10}$~erg~cm$^{-2}$~s$^{-1}$.

We have compared the physical properties of \lmc\ with other
supergiant shells, superbubbles, and SNRs in the LMC.  The
derived electron densities of the \lmc\ regions are generally 
comparable to the values determined for the supergiant shell 
LMC\,4 and the superbubble N\,11, but are lower than those 
determined for the superbubble N\,44 and a sample of SNRs 
in the LMC.  The plasma temperature of \lmc\ is comparable
to that determined for LMC\,4 and N\,11, but is generally lower 
than the plasma temperatures determined N\,44, and SNRs in 
the LMC.

\acknowledgments
{This work is supported by NASA grants NAG 5-2973 and NAG 5-8104.
This research has made use of data obtained from the High Energy 
Astrophysics Science Archive Research Center (HEASARC), provided 
by NASA's Goddard Space Flight Center.  SDP wishes to thank J. 
Turner for many helpful discussions on the reduction of \rosat\ 
PSPC data using XSPEC.  SDP also wishes to thank Q. D. Wang and
for his helpful comments.}

\appendix

\section{\rosat\ Pointed Observations Toward \lmc}

Here we present Tables~A1 and A2 which list the individual \rosat\ 
pointed observations that were mosaicked to produce the PSPC and HRI 
mosaics, respectively.  These tables contain the number of data sets
for each observation, the pointing direction, exposure time, and 
target name.

\placetable{tbl:pspc-obs}

\placetable{tbl:hri-obs}

\clearpage

\begin{center} {\large \bf Figure Captions} \end{center}

\figcaption {(a) \rosat\ PSPC mosaic of the \lmc\ region in the R4-R7
bandpass with the areas from which spectra were extracted labelled. (b)
\rosat\ HRI mosaic of the \lmc\ region with X-ray sources mentioned in
the text labeled. \label{fig:fig1ab}}

\figcaption {(a) PDS \ha\ image of \lmc\ and the surrounding region with
PSPC contours overlaid (Courtesy R. Kennicutt).  The contour levels are 
in increments of 2e-3, starting at 2e-3 and ending at 2e-2.  (b) Same as 
(a) with \hii\ regions and SNRs mentioned in the text labeled. 
\label{fig:fig2ab}}

\figcaption {ATCA map of \lmc\ integrated over the velocity range of 
$V_{hel} = 190 - 387$~\kms\ with PSPC contours overlaid.  The contour 
levels are the same as in Figure~2\,a.  The peak temperature of the 
plotted \hi\ map is 99~K.\label{fig:fig3}}

\figcaption {X-ray spectra and floating $N_H$ model fits for the 
(a) PSPC-North, (b) PSPC-Arc, (c) PSPC-Spur, (d) SIS~0 Chip~1, and 
(e) SIS~1 Chip~3 regions.  Both (d) and (e) were fit simultaneously 
with the model, but are shown here as separate plots for clarity.
\label{fig:fig4}}

\figcaption {$\chi^2$ grid plots for the (a) PSPC-North, (b) PSPC-Arc, 
(c) PSPC-Spur, and (d) SIS~0 Chip~1 and SIS~1 Chip~3 combined spectral 
fit.  The three contours correspond to 68\%, 90\%, and 99\% confidence 
levels.  Temperatures are given in keV.\label{fig:fig5}}

\clearpage

\begin{deluxetable}{lcc}
\footnotesize
\tablecaption{Broad Energy Band Definitions. \label{tbl:pspc-bands}}
\tablewidth{0pt}
\tablehead{
\colhead{Band Name} & \colhead{PI Channels}
    & \colhead{Energy\tablenotemark{a}~~(keV)}}
\startdata
R1              & $8-19$    & $0.11-0.284$ \nl
R2              & $20-41$   & $0.14-0.284$ \nl
R4              & $52-69$   & $0.44-1.01$  \nl
R5              & $70-90$   & $0.56-1.21$  \nl
R6              & $91-131$  & $0.73-1.56$  \nl
R7              & $132-201$ & $1.05-2.04$  \nl
\enddata
\tablenotetext{a} {\footnotesize 10\% of peak response.}
\end{deluxetable}

\clearpage

\begin{deluxetable}{lcccc}
\footnotesize
\tablecaption{Relevant Parameters of the X-ray Mosaics. \label{tbl:stats}}
\tablewidth{0pt}
\tablehead{
\colhead{Band} & \colhead{Average} & \colhead{Total Counts}
    & \colhead{Total Counts} & \colhead{Average} \\
    & \colhead{Exposure (ksec)} & \colhead{Observed}
    & \colhead{Background}   & \colhead{Intensity\tablenotemark{a}}}
\startdata
R4  & 25.12  & 289179  & 46063   & 255   \nl
R5  & 25.15  & 442372  & 19820   & 389   \nl
R6  & 24.65  & 655163  & 12646   & 410   \nl
R7  & 21.94  & 419176  & 13250   & 179   \nl
HRI & 26.13  & 3149624 & 2346460 & 1032  \nl
\enddata
\tablenotetext {a}{\footnotesize Units of $10^{-6}$~\csam.}
\end{deluxetable}

\clearpage

\begin{deluxetable}{lcccccc}
\footnotesize
\tablecaption{Parameters of the Extraction Regions. \label{tbl:reg-prop}}
\tablewidth{0pt}
\tablehead{
\colhead{Region} & \colhead{RA$_{J2000}$} & \colhead{Dec$_{J2000}$} & 
\colhead{Size} & \colhead{Offaxis} & \colhead{Net} & \colhead{Average} \\
\colhead{Name}   & \colhead{(h m s)}      & \colhead{(\arcdeg~\arcmin~
\arcsec)} & \colhead{($\Box '$)} & \colhead{Angle ($'$)} 
& \colhead{Counts}  & \colhead{Exposure (ksec)}}
\startdata
PSPC-North   & 5~44~19 & -69~10~51 & 185    & 34.7  &  717   &  2.7   \nl
PSPC-Arc     & 5~41~55 & -69~34~23 & 166.5  & 27.8  & 1255   &  2.7   \nl
PSPC-Spur    & 5~44~02 & -70~12~46 & 123    & 34.2  &  512   &  2.8   \nl
SIS~0 Chip~1 & 5~46~04 & -69~13~03 & 108    &  8.4  & 3737   & 30.7   \nl
SIS~1 Chip~3 & 5~46~04 & -69~13~03 & 104    &  8.1  & 3382   & 30.7   \nl
\enddata
\end{deluxetable}

\clearpage

\begin{deluxetable}{lcccc}
\footnotesize
\tablecaption{Observed X-ray Properties Toward \lmc \label{tbl:xray-fit}}
\tablewidth{0pt}
\tablehead{
\colhead{Region} & \colhead{PSPC-North} & \colhead{PSPC-Arc} & 
\colhead{PSPC-Spur} & \colhead{SIS\tablenotemark{a}} \\
\cline{1-5} \\
\multicolumn{5}{c}{Free $N_H$ Fits}} 
\startdata
kT (keV)\tablenotemark{b}                     & 0.14  & 0.23 & 0.17 & 0.32 \nl
$N_H$ (10$^{22}$~cm$^{-2}$)$^{\rm b}$         & 0.87  & 0.67 & 0.62 & 0.32 \nl
kT (keV)\tablenotemark{c}                     & 0.11 -- 0.28 & 0.13 -- 0.32 & 
0.07 -- 0.37 & 0.28 -- 0.37; 0.60 -- 0.69 \nl
$N_H$ (10$^{22}$~cm$^{-2}$)$^{\rm c}$         & 0.58 -- 1.08 & 0.48 -- 0.94 & 
0.12 -- 1.43 & 0.24 -- 0.42; 0.19 -- 0.24 \nl
\cutinhead{Fixed $N_H$ Fits}
kT (keV)\tablenotemark{b}                     & 0.31  & 0.33 & 0.23 & 0.29  \nl
$N_H$ (10$^{22}$~cm$^{-2}$)\tablenotemark{d}  & 0.48  & 0.43 & 0.42 & 0.47  \nl
kT (keV)\tablenotemark{c}                     & 0.09 -- 0.67 & 0.10 -- 0.47 &
0.08 -- 0.45 & 0.27 -- 0.33; 0.60 -- 0.69 \nl
$N_H$ (10$^{22}$~cm$^{-2}$)$^{\rm c}$         & 0.08 -- 1.25 & 0.28 -- 1.05 &
0.07 -- 1.35 & 0.34 -- 0.54; 0.12 -- 0.17 \nl
\enddata
\tablenotetext{a}{Combined model fit of the SIS~0 Chip~1 \& SIS~1 
Chip~3 spectra.}
\tablenotetext{b}{Parameter value is the best fit value of the model.}
\tablenotetext{c}{Parameter range gives 90\% confidence levels.}
\tablenotetext{d}{Parameter value is the sum of the Galactic and LMC
components of $N_H$ toward \lmc\ as given by Dickey \& Lockman (1990) 
and Staveley-Smith (2000), respectively.}
\end{deluxetable}

\clearpage

\begin{deluxetable}{lccccccc}
\tablecolumns{7}
\footnotesize
\tablecaption{Physical Properties of Hot Gas in the Extracted 
Regions\tablenotemark{a}\label{tbl:xray-prop}}
\tablewidth{0pt}
\tablehead{
\colhead{Region} & \colhead{$l$} & \colhead{$F_X$\tablenotemark{b}} & 
\colhead{L$_X$\tablenotemark{b,c}} & \colhead{$\Lambda$\tablenotemark{b}} & 
\multicolumn{2}{c}{n$_e$\tablenotemark{d}} \\
\colhead{Name}  & \colhead{(pc)} & \colhead{(10$^{-11}$ 
erg cm$^{-2}$ s$^{-1}$)} & \colhead{(10$^{36}$ erg s$^{-1}$)} & 
\colhead{($10^{-24}$ erg cm$^{3}$ s$^{-1}$)} & 
\multicolumn{2}{c}{(cm$^{-3}$)} \\
\cline{6-7} \\
\colhead{} & \colhead{} & \colhead{} & \colhead{} & \colhead{} & 
\colhead{$f= 0.5$} & \colhead{$f = 1.0$}  } 
\startdata
%NAME                 &  l  &  Fx   &  Lx    & lam  &     ne
%                                                      0.5    1.0          
PSPC-North            & 140 & 1.67  &  5.0   & 7.43 & 0.09  & 0.07   \nl
PSPC-Arc              & 230 & 1.92  &  5.7   & 7.75 & 0.08  & 0.06   \nl
PSPC-Spur             & 400 & 1.65  &  4.9   & 5.76 & 0.08  & 0.05   \nl
SIS\tablenotemark{e}  & 140 & 0.68  &  2.0   & 7.13 & 0.08  & 0.06   \nl
\enddata
\tablenotetext{a}{Determined using the fixed $N_H$ X-ray model fits given
in Table~4.}
\tablenotetext{b}{Measured in the 0.44 -- 2.04 keV range (R4-R7 band).}
\tablenotetext{c}{Assuming an LMC distance of 50~kpc.}
\tablenotetext{d}{Derived using the assumed path length, $l$, and the
projected surface area from Table~3.}
\tablenotetext{e}{Derived using the combined X-ray model fits to the SIS
data.} 
\end{deluxetable}

\clearpage

\begin{deluxetable}{lcccc}
\tablecolumns{4}
\footnotesize
\tablecaption{Derived Physical Conditions of the Hot Gas\tablenotemark{a} 
\label{tbl:xray-para}}
\tablewidth{0pt}
\tablehead{
\colhead{Region} & \colhead{$n_e$}       & \colhead{$\Sigma_M$}    & 
\colhead{$\Sigma_{E_{th}}$} & \colhead{$N_{SNe}$}\\
\colhead{} & \colhead{(cm$^{-3}$)} & \colhead{($M_{\odot}$)} &
\colhead{($10^{50}$~erg~arcmin$^{-2}$)} & \colhead{} }
\startdata
PSPC-North          & 0.07 -- 0.09 &  46 --  65 & 0.32 -- 0.46 &  80 -- 115 \nl
PSPC-Arc            & 0.06 -- 0.08 &  65 --  92 & 0.48 -- 0.68 & 120 -- 170 \nl
PSPC-Spur           & 0.05 -- 0.08 & 107 -- 152 & 0.56 -- 0.79 & 140 -- 198 \nl
SIS\tablenotemark{b}& 0.06 -- 0.08 &  41 --  58 & 0.27 -- 0.38 &  68 --  95 \nl
\enddata
\tablenotetext{a}{Results derived using the fixed $N_H$ model X-ray
spectral fits.}
\tablenotetext{b}{Results from the combined SIS model fits.}
\end{deluxetable}

\clearpage

\begin{deluxetable}{lccccl}
\footnotesize
\tablecaption{Physical Conditions of Hot Gas in Other Objects 
in the LMC\label{tbl:xray-other}}
\tablewidth{0pt}
\tablehead{
\colhead{Object ID} & \colhead{$kT$} & \colhead{$F_X$} 
& \colhead{$n_e$} & \colhead{$\Sigma_{E_{th}}$}  & \colhead{Reference}  \\
\colhead{}          & \colhead{(keV)} & 
\colhead{(10$^{-10}$~erg~cm$^{-2}$~s$^{-1}$)} & \colhead{(cm$^{-3}$)} 
& \colhead{($10^{50}$~erg~arcmin$^{-2}$)} & \colhead{} } 
\startdata
LMC\,2     & 0.43 & 0.4\tablenotemark{a} & 0.02\tablenotemark{b}  &
0.46\tablenotemark{b}  & Wang \& Helfand (1991)                \nl
LMC\,2     & 0.43 & 0.6\tablenotemark{a} & 0.009\tablenotemark{c} &
0.41\tablenotemark{c}  & Wang \& Helfand (1991)                 \nl
LMC\,4     & 0.21 & 0.5                  & 0.008\tablenotemark{d} & 
0.32\tablenotemark{d}   & Bomans, Dennerl, \&  K\"urster (1994)  \nl
N\,11      & 0.3        & 0.01           & 0.06\tablenotemark{e}  &
0.23\tablenotemark{e}   & Mac Low et al. (1998)                  \nl
N\,44      & 0.55       & 0.02           & 0.14\tablenotemark{f}  &
0.07\tablenotemark{f}   & Magnier et al. (1996)                  \nl
DEM~316\,A & 0.8  & 0.009                & 0.28                   &
0.16                    & Williams et al. (1997)                 \nl
DEM~316\,B & 0.9  & 0.007                & 0.15                   &
0.08                    & Williams et al. (1997)                 \nl
N\,11\,L   & 0.3  & 0.004                & 0.85                   &
0.13                    & Williams et al. (1999)                 \nl
N\,86      & 0.15 & 0.008                & 0.16                   &
0.03                    & Williams et al. (1999)                 \nl
\enddata
\tablenotetext{a}{Measured in the \einst\ bandpass of 0.4 -- 4.0~keV.}
\tablenotetext{b}{Derived using a radius of 250~pc covering 
the bright X-ray arc.}
\tablenotetext{c}{Derived using a radius of 500~pc for the 
entire supergiant shell.}
\tablenotetext{d}{Derived assuming that LMC\,4 is cylindrical
in shape with a radius of 600~pc and a height of 1200~pc.}
\tablenotetext{e}{Derived assuming a radius of 60~pc for 
the superbubble and a hot gas emissivity of $\lambda = 
3.5 \times 10^{-24}$erg~cm$^3$~s$^{-1}$.}
\tablenotetext{f}{Derived assuming a radius of 45~pc and a 
hot gas filling factor of $f = 0.5$.}
\end{deluxetable}

\clearpage

\begin{deluxetable}{lccccl}
\tablenum{A1}
\footnotesize
\tablecaption{PSPC observations included in mosaic. \label{tbl:pspc-obs}}
\tablewidth{0pt}
\tablehead{
\colhead{Sequence} & \colhead{Sets\tablenotemark{a}} & \colhead{RA$_{2000}$}
    & \colhead{Dec$_{2000}$} & \colhead{Exposure (ksec)} & \colhead{Target}}
\startdata
500138 & 2 & 5 26 36 & -68 50 23 & 27.08  &  N144             \nl
400148 & 1 & 5 27 47 & -69 54 00 &  5.90  &  RX J0527.8-6954  \nl
400298 & 2 & 5 27 47 & -69 54 00 & 14.85  &  RX J0527.8-6954  \nl
300172 & 3 & 5 32 28 & -70 21 36 & 12.54  &  NOVA LMC 88 \#1  \nl
500100 & 2 & 5 35 28 & -69 16 11 & 24.84  &  SN1987A          \nl
500303 & 1 & 5 35 28 & -69 16 11 &  9.16  &  SN 1987 A        \nl
600100 & 1 & 5 35 38 & -69 16 11 & 12.16  &  REGION F         \nl
300335 & 1 & 5 36 12 & -70 45 00 &  9.90  &  2 NEW SUPERSOFT SRCS  \nl
500131 & 1 & 5 38 33 & -69 06 36 & 15.45  &  N157             \nl
400079 & 1 & 5 39 38 & -69 44 23 &  5.20  &  LMC X-1          \nl
150044 & 1 & 5 40 12 & -69 19 48 &  5.15  &  PSR 0540-69      \nl
400052 & 1 & 5 40 12 & -69 19 48 &  6.55  &  PSR 0540-69      \nl
400133 & 1 & 5 40 12 & -69 19 48 &  1.72  &  PSR 0540-69      \nl
400012 & 1 & 5 46 45 & -71 09 00 & 14.92  &  CAL87            \nl
400013 & 1 & 5 46 45 & -71 09 00 & 14.73  &  CAL87            \nl
500259 & 1 & 5 47 09 & -69 41 59 &  3.92  &  DEM 316          \nl
\enddata
\tablenotetext{a}{Number of separate data sets processed for observation.}
\end{deluxetable}

\clearpage

\begin{deluxetable}{lccccl}
\tablenum{A2}
\footnotesize
\tablecaption{HRI observations included in mosaic. \label{tbl:hri-obs}}
\tablewidth{0pt}
\tablehead{
\colhead{Sequence} & \colhead{Sets\tablenotemark{a}} & \colhead{RA$_{2000}$}
    & \colhead{Dec$_{2000}$} & \colhead{Exposure (ksec)} & \colhead{Target}}
\startdata
400124 & 1 & 5 18 40.8 & -68 14 24 &   4.55 &  LMC H 204        \nl
500171 & 3 & 5 19 33.6 & -69 02 24 &  17.32 &  0519.69.0        \nl 
400657 & 1 & 5 20 31.2 & -69 31 48 &   3.83 &  RXJ0520.5-6932   \nl
400153 & 2 & 5 22 26.4 & -67 58 12 &   8.56 &  N44C/STAR 2      \nl
400231 & 1 & 5 22 28.8 & -69 26 24 &   3.89 &  LMC FIELD 1      \nl
600036 & 1 & 5 22 33.6 & -69 18 36 &   1.13 &  LMC H 20         \nl
500002 & 1 & 5 25 2.4  & -69 38 24 &  27.25 &  N132D            \nl
600032 & 1 & 5 25 43.2 & -69 13 12 &   0.71 &  CAL 37           \nl
600031 & 1 & 5 25 52.8 & -70 11 24 &   1.26 &  CAL38            \nl
400353 & 2 & 5 26 9.6  & -67 42 36 &   7.94 &  LMC FIELD A      \nl
400238 & 1 & 5 26 28.8 & -69 46 48 &   3.99 &  H211             \nl
600033 & 1 & 5 27 38.4 & -69 10 48 &   1.65 &  CAL 40           \nl
201689 & 1 & 5 27 48   & -69 54 00 &   8.50 &  RX J 0527.8-6954 \nl
600647 & 1 & 5 28 19.2 & -68 30 36 &  23.91 &  LMC POINT 017    \nl
600646 & 1 & 5 28 19.2 & -68 51 00 &  24.01 &  LMC POINT 016    \nl
600641 & 1 & 5 28 19.2 & -69 10 48 &  26.71 &  LMC POINT 011    \nl
600782 & 1 & 5 28 19.2 & -69 51 00 &  12.79 &  LMC POINT AO6-10 \nl
400356 & 1 & 5 28 33.6 & -68 36 36 &   3.16 &  W528.8-684       \nl
400355 & 1 & 5 28 43.2 & -67 25 48 &   4.60 &  W529.8-672       \nl
600777 & 1 & 5 28 50.4 & -69 33 00 &  19.84 &  LMC POINT AO6-05 \nl
600645 & 1 & 5 32 4.8  & -68 30 36 &  19.00 &  LMC POINT 015    \nl
600640 & 1 & 5 32 4.8  & -68 51 00 &  23.98 &  LMC POINT 010    \nl
600635 & 1 & 5 32 4.8  & -69 10 48 &  21.23 &  LMC POINT 005    \nl
600655 & 1 & 5 32 4.8  & -70 10 48 &  11.99 &  LMC POINT 026    \nl
400125 & 1 & 5 32 7.2  & -69 19 12 &   4.92 &  LMC H 215        \nl
400660 & 1 & 5 32 16.8 & -71 40 12 &   5.52 &  RXJ0532.3-7140   \nl
400234 & 2 & 5 32 52.8 & -67 43 12 &   6.39 &  LMC FIELD 2      \nl
500234 & 1 & 5 33 60   & -69 55 12 &   7.66 &  0534-69.9        \nl
400352 & 2 & 5 35 4.8  & -70 39 00 &   5.43 &  LMC FIELD B      \nl
400056 & 1 & 5 35 28.8 & -69 16 12 &  23.47 &  SN 1987A         \nl
600634 & 1 & 5 35 52.8 & -68 51 00 &  23.62 &  LMC POINT 004    \nl
600650 & 1 & 5 35 52.8 & -69 51 00 &   2.10 &  LMC POINT 020    \nl
400230 & 1 & 5 35 55.2 & -69 21 00 &   4.86 &  LMC FIELD 3      \nl
400233 & 1 & 5 35 55.2 & -69 56 24 &   4.98 &  LMC FIELD 4      \nl
600035 & 1 & 5 36 2.4  & -67 34 48 &   0.78 &  CAL 60           \nl
600639 & 1 & 5 36 7.2  & -68 31 48 &  25.62 &  LMC POINT 009    \nl
400644 & 1 & 5 36 16.8 & -67 18 00 &   4.57 &  RXJ0536.3-6718   \nl
500036 & 1 & 5 37 52.8 & -69 10 12 &  10.91 &  N 157B           \nl
400642 & 1 & 5 38 16.8 & -69 24 00 &   2.43 &  CAL 69           \nl
600228 & 1 & 5 38 43.2 & -69 06 00 &  30.09 &  30 DOR           \nl
400649 & 1 & 5 38 52.8 & -69 01 48 &   2.72 &  CAL 74           \nl
600638 & 1 & 5 39 38.4 & -68 30 36 &  25.73 &  LMC POINT 008    \nl
600633 & 1 & 5 39 38.4 & -68 51 00 &  21.50 &  LMC POINT 003    \nl
150013 & 1 & 5 39 38.4 & -69 44 24 &   0.80 &  LMC X-1          \nl
400650 & 1 & 5 39 43.2 & -69 45 00 &   1.20 &  LMC X-1          \nl
600774 & 1 & 5 40 4.8  & -69 31 48 &  22.34 &  LMC POINT AO6-02 \nl
150008 & 1 & 5 40 12   & -69 19 48 &  21.59 &  SNR0540-69.3     \nl
400236 & 1 & 5 40 45.6 & -70 12 00 &   1.55 &  LMC FIELD 6      \nl
400349 & 2 & 5 41 26.4 & -68 49 48 &   6.56 &  LMC FIELD C      \nl
500235 & 1 & 5 43 7.2  & -68 58 48 &  28.24 &  0543-68.9        \nl
600643 & 1 & 5 43 26.4 & -68 30 36 &  21.45 &  LMC POINT 013    \nl
600632 & 1 & 5 43 26.4 & -68 51 00 &  22.49 &  LMC POINT 007    \nl
600637 & 1 & 5 43 26.4 & -68 51 00 &  22.49 &  LMC POINT 007    \nl
600773 & 1 & 5 43 26.4 & -69 30 36 &  15.99 &  LMC POINT AO6-01 \nl
600648 & 1 & 5 43 26.4 & -69 51 00 &   1.50 &  LMC POINT 018    \nl
600778 & 1 & 5 43 26.4 & -69 51 00 &   8.79 &  LMC POINT AO6-06 \nl
400121 & 1 & 5 46 28.8 & -68 34 12 &   1.59 &  CAL 86           \nl
400123 & 1 & 5 46 57.6 & -68 52 12 &   1.39 &  LMC H 232        \nl
500232 & 2 & 5 47 9.6  & -69 42 00 &  20.35 &  0547-69.7        \nl
600644 & 1 & 5 47 12   & -68 30 36 &  18.76 &  LMC POINT 014    \nl
600642 & 1 & 5 47 12   & -68 51 00 &  21.10 &  LMC POINT 012    \nl
600636 & 1 & 5 47 12   & -69 10 48 &  20.63 &  LMC POINT 006    \nl
600631 & 1 & 5 47 12   & -69 30 36 &  25.65 &  LMC POINT 001    \nl
500233 & 2 & 5 47 50.4 & -70 24 36 &  26.41 &  0548-70.4        \nl
400659 & 1 & 5 47 52.8 & -68 22 48 &   5.29 &  RXJ0547.9-6823   \nl
\enddata
\tablenotetext{a}{\footnotesize Number of separate data sets processed 
for observation.}
\end{deluxetable}


\clearpage
 
\begin{thebibliography}{}
\bibitem[Bessell 1991]{bessell91} Bessell, M. S. 1991, A\&A, 242, L17
\bibitem[Bomans et al. 1994]{bdk94} Bomans, D. J., Dennerl, K., \& 
     K\"urster, M. 1994, A\&A, 283, L21
\bibitem[Chu \& Mac Low 1990]{cm90} Chu, Y.-H., \& Mac Low, M.-M. 1990
     ApJ, 365, 510
\bibitem[Chu \& Snowden 1996]{cs96} Chu, Y.-H., \& Snowden, S. L. 1996, 
     {\it ROSAT} Newsletter \#13, ed. S. L. Snowden (NASA:GSFC), 26
\bibitem[Dickey \& Lockman 1990]{dl90} Dickey, J. M., \& Lockman, F. J.
     1990, ARA\&A, 28, 215
\bibitem[Feast 1999]{feast99} Feast, M. 1999, PASP, 111, 775
\bibitem[Fukui et al. 1999]{fukui99} Fukui, Y., et al. 1999, PASJ, 51, 745
\bibitem[Goudis \& Meaburn 1978]{gm78} Goudis, C., \& Meaburn, J. 1978,
     A\&A, 68, 189
\bibitem[Greisen \& Calabretta 1996]{gre96} Greisen, E. W., \& Calabretta,
     M. 1995, ASP Conf. Ser. 77, Astronomical Data Analysis Software
     and Systems IV, ed. R. A. Shaw, H. E. Payne, \& J. J. E. Hays
     (San Francisco: ASP), 223
\bibitem[Henize 1956]{hen56} Henize, K. G. 1956, ApJS, 2, 315
\bibitem[Hughes et al. 1998]{h98} Hughes, J. P., Hayashi, I., \& Koyama, 
     K. 1998, ApJ, 505, 732
\bibitem[Kennicutt \& Hodge 1986]{kh86} Kennicutt, R. C., \& Hodge, P. W.
     1986, ApJ, 306, 130
\bibitem[Kuntz \& Snowden]{esas2} Kuntz, K. D., \& Snowden, S. L. 1998,
     Cookbook for Analysis Procedures for {\it ROSAT} XRT Observations of 
     Extended Objects and the Diffuse Background, Part~II: Mosaics,
     US \rosat\ Science Data Center, NASA/GSFC
\bibitem[Mac Low et al. 1998]{mmml98} Mac Low, M.-M., Chang, T. H., Chu, 
     Y.-H., Points, S. D., Smith, R. C., \& Wakker, B. P. 1998, ApJ, 493, 260
\bibitem[Magnier et al. 1996]{gm96} Magnier, E. A., Chu, Y.-H., Points, S. D.,
     Hwang, U., \& Smith, R. C. 1996, ApJ, 464, 829
\bibitem[McKee \& Ostriker 1977]{mo77} McKee, C. F., \& Ostriker, J. P. 1977,
     ApJ, 218, 148
\bibitem[Meaburn 1980]{m80} Meaburn, J. 1980, MNRAS, 192, 365
\bibitem[Mochizuki et al. 1994]{mochi94} Mochizuki, K., Nakagawa, T., Doi,
     Y., Yui, Y. Y., Okuda, H., Shibai, H., Yui, M., Nishimura, T., \&
     Low, F. J. 1994, ApJ, 430, L37 
\bibitem[Morrison \& McCammon]{mm83} Morrison, R., \& McCammon, D.
     1983, ApJ, 270, 119
\bibitem[Points et al. 1999]{sdp99} Points, S. D., Chu, Y.-H., Kim, S., 
     Smith, R. C., Snowden, S. L., Brandner, W., \& Gruendl, R. A. 1999,
     ApJ, 518, 298 (Paper~I)
\bibitem[Raymond \& Smith 1977]{rs77} Raymond, J. C., \& Smith, B. W.
     1977, ApJS, 35, 419
\bibitem[Schlegel et al. 1994]{s94} Schlegel, E. M., Marshall, F. E.,
     Mushotzky, R. F., Smale, A. P., Weaver, K. A., Serlemitsos, P. J.,
     Petre, R., \& Jahoda, K. M. 1994, ApJ, 422, 243
\bibitem[Slavin \& Cox 1992]{sc92} Slavin, J. D., \& Cox, D. P. 1992, ApJ,
     392, 131
\bibitem[Snowden \& Kuntz 1998]{esas1} Snowden, S. L., \& Kuntz, K. D. 1998,
     Cookbook for Analysis Procedures for \rosat\ XRT Observations of
     Extended Objects and the Diffuse Background, Part~1: Individual
     Observations, US \rosat\ Science Data Center, NASA/GSFC
\bibitem[Snowden et al. 1994]{smbm94} Snowden, S. L., McCammon, D.,
     Burrows, D. N., \& Mendenhall, J. A. 1994, \apj, 424, 714
\bibitem[Snowden \& Petre 1994]{sp94} Snowden, S. L., \& Petre, R. 1994,
     ApJL, 436, L123
\bibitem[Staveley-Smith 2000]{lss00} Staveley-Smith, L. 2000, private
     communication
\bibitem[Tanaka et al. 1994]{tih94} Tanaka, Y., Inoue, H., \& Holt, S. S.
     1994, PASJ, 46, L37
\bibitem[Tenorio-Tagle \& Bodenheimer 1988]{ttb88} Tenorio-Tagle, G., \&
     Bodenheimer, P. 1988, ARA\&A, 26, 145
\bibitem[Wang 1999]{wang99} Wang, Q. D. 1999, ApJ, 510, L139
\bibitem[Wang \& Helfand 1991]{wh91} Wang, Q. D., \& Helfand, D. J. 1991,
     ApJ, 379, 327
\bibitem[Williams et al. 1997]{rosa97} Williams, R. M., Chu, Y.-H., 
     Dickel, J. D., Beyer, R., Petre, R., Smith, R. C., \& Milne, D. K.
     1997, ApJ, 480, 618
\bibitem[Williams 1999]{rosathesis} Willaims, R. M. 1999, Ph.D. Thesis, 
     University of Illinois at Urbana-Champaign
\bibitem[Williams et al. 1999]{rosa99} Williams, R. M., Chu, Y.-H.,
     Dickel, J. D., Smith, R. C., Milne, D. K., \& Winkler, P. F. 1999
     ApJ, 514, 798
\end{thebibliography}
\end{document}